# A Meta-learning based Distribution System Load Forecasting Model Selection Framework

Yiyan Li[*], Si Zhang, Rongxing Hu, Ning Lu

*Electrical & Computer Engineering Department, Future Renewable Electric Energy Delivery and Management (FREEDM) Systems Center, North Carolina State University, Raleigh, NC 27606 USA*

**Abstract:** This paper presents a meta-learning based, automatic distribution system load forecasting model selection framework. The framework includes the following processes: feature extraction, candidate model preparation and labeling, offline training, and online model recommendation. Using load forecasting needs and data characteristics as input features, multiple metalearners are used to rank the candidate load forecast models based on their forecasting accuracy. Then, a scoring-voting mechanism is proposed to weights recommendations from each meta-leaner and make the final recommendations. Heterogeneous load forecasting tasks with different temporal and technical requirements at different load aggregation levels are set up to train, validate, and test the performance of the proposed framework. Simulation results demonstrate that the performance of the meta-learning based approach is satisfactory in both seen and unseen forecasting tasks.

| *Scalar* | | $w$ | One set of the Metalearner parameters |
|---|---|---|---|
| $C$ | Binary variable in Similar Day (SD) model | $w^*$ | Optimal set of the Metalearner parameters |
| $D$ | Number of total features | $\beta_1, \beta_2, \beta_3$ | Decay factors of the SD model |
| $i_B$ | Index of the LF model | $\theta$ | One set of the LF model parameters |
| $I_B$ | Number of LF models | $\theta^*$ | Optimal set of the LF model parameters |
| $i_M$ | Index of the metalearner | $\gamma$ | Similarity between different days |
| $I_M$ | Number of metalearners | $\eta$ | Accuracy of the metalearner |
| $j$ | Index of the LF task | *Matrix/vector* | |
| $J$ | Number of LF tasks | $\mathbf{X}_j$ | Input data for the $j^{th}$ LF task |
| $K_j$ | Training data length of the $j^{th}$ LF task | $\mathbf{X}_j^{train}$ | Training data for the $j^{th}$ LF task |
| $K_M$ | Number of LF tasks for metalearner training | $\mathbf{X}_j^{test}$ | Testing data for the $j^{th}$ LF task |
| $m$ | Index of the attribute | $\mathbf{y}_j$ | Actual load for the $j^{th}$ LF task |
| $M_j$ | Number of attributes for the $j^{th}$ LF task | $\mathbf{y}_j^{train}$ | Actual load for training in the $j^{th}$ LF task |
| $n$ | Index of the input data | $\mathbf{y}_j^{test}$ | Actual load for testing in the $j^{th}$ LF task |
| $N_j$ | Number of the input data for the $j^{th}$ LF task | $\hat{\mathbf{y}}_{j,i_B}^{train}$ | Forecasted load by LF model $i_B$ on $\mathbf{X}_j^{train}$ |
| $p$ | Order of the AR part of the ARIMA model | $\hat{\mathbf{y}}_{j,i_B}^{test}$ | Forecasted load by LF model $i_B$ on $\mathbf{X}_j^{test}$ |
| $q$ | Order of the MA part of the ARIMA model | $\hat{\mathbf{y}}^{new}$ | Forecasted load for new LF tasks |
| $R$ | Number of LF tasks requirement | $\mathbf{F}$ | Features of a LF Task |
| $S$ | Scores of the candidate LF model | $\mathbf{F}^{train}$ | Input features to train the metalearner |
| $T_{fore}$ | Index of forecasting day | $\mathbf{F}^{test}$ | Input features to test the metalearner |

* Corresponding author. *Tel.:* +1 919-903-1199; *E-mail address:* yli257@ncsu.edu.

| | | | |
|---|---|---|---|
| $T_{hist}$ | *Index of historical day* | $\mathbf{F}^{new}$ | *Features of a new LF task* |
| $\Delta T$ | *Distance between $T_{fore}$ and $T_{hist}$* | $\Phi$ | *Actual best model* |
| $\Phi^{train}$ | *Actual best models to train the metalearner* | $\mathbf{Z}(i_B, j)$ | *RMSE of LF model $i_B$ in the $j^{th}$ LF task* |
| $\Phi^{test}$ | *Actual best models to test the metalearner* | ***Functions*** | |
| $\hat{\Phi}_{i_M}^{train}$ | *Recommended models by $i_M$ on $\mathbf{F}^{train}$* | $f_\theta^{i_B}$ | *LF model $i_B$ with parameter $\theta$* |
| $\hat{\Phi}_{i_M}^{test}$ | *Recommended models by $i_M$ on $\mathbf{F}^{test}$* | $g_w^{i_M}$ | *Metalearner $i_M$ with parameter $w$* |
| $\hat{\Phi}_{i_M}^{new}$ | *Recommended models by $i_M$ on $\mathbf{F}^{new}$* | $h_{i_M}$ | *Score-accuracy function of metalearner $i_M$* |
| $\hat{\Phi}^{train}$ | *Recommended model after voting on $\mathbf{F}^{train}$* | $\mathcal{L}^{base}$ | *Loss function of the LF model* |
| $\hat{\Phi}^{test}$ | *Recommended model after voting on $\mathbf{F}^{test}$* | $\mathcal{L}^{meta}$ | *Loss function of the metalearner* |
| $\hat{\Phi}^{new}$ | *Recommended model after voting on $\mathbf{F}^{new}$* | | |

## 1. Introduction

The needs for load forecasting (LF) have increased drastically at all levels in power distribution systems accompanied with the increasing penetration of the distributed generation resource (DER) [1]. In recent years, many LF models have been developed to solve distribution system LF problems, including but not limited to deep-learning based models [2]-[4], probabilistic-forecasting models [5]-[7] and hybrid-forecasting models [8]-[9]. Different ensemble learning methods [10]-[12] have also been investigated to further improve LF forecasting accuracy. Thus, for a given LF task, an optimal LF model can usually be found by comparing all exsiting methods. However, at different aggregation levels in a power distribution system, LF tasks for different applications may have very different temporal and technical requirements to meet. The integration of DERs, microgrids, and demand response programs also makes LF tasks an evolving process. Consequently, the need for selecting an optimal LF model is more frequent. However, very few research efforts have been focused on developing automated, credible, and robust LF model selection, the objective of which is to select the best LF model (or a few suitable LF models) for a given LF task specified by the characteristic available data sets and LF requirements on a timely manner.

Traditionally, the Knowledge-based expert system (KES) approach is used for selecting forecasting models [13]-[16]. The main disadvantage of the KES approach is inflexibility. Whenever new models are introduced or new forecasting scenarios are considered, a manual update of the system rules is required, making the maintenance costs high. Moreover, KES cannot be used for unseen LF tasks. Thus, the KES approach is inadequate for selecting an LF model in an active distribution network (ADN), where LF tasks are heterogeneous in terms of scale, input data characteristics, and LF requirements.

In recent years, meta-learning [17]-[21], generally interpreted as 'learning to learn', is introduced to provide model recommendations for different machine learning tasks. In [22], Cui et al. propose to recommend proper LF models for different type of buildings, where the Artificial Neural Network (ANN) is selected to construct the mapping from features to LF models. Also focusing on the building-level LF problem, the authors in [23] include user preferences to further enhance the model recommendation efficacy. In [24], Matijas et al. prepare 7 candidate models to deal with 65 forecasting tasks where statistical features are created to quantify these tasks and classical classifiers are applied to construct the mapping from task features to optimal models. In [25], Arjmand et al. test a similar system with 6 candidate models and 18 features on 30 forecasting tasks generated from zonal data of Ontario, Canada, where ReliefF is used to assist feature selection and improve the accuracy of model recommendation. In [26], Wang et al. consider both rule-based and meta-learning methods to support the forecasting model selection for univariate time series, where self-organization Map (SOM) is introduced to create and visualize forecasting tasks. Also focusing on univariate time series forecasting, Talagala et al. proposed a similar work in [27] using a larger feature set on more candidate models and tests conducted on monthly, quarterly and yearly time series are considered as different forecasting tasks. In [28], Lemke and Gabrys combine several forecasting methods based on the ranking results provided by the meta-learning system to enhance the forecasting stability. In [29], Heng et al. introduce the framework of meta-learning to wind power forecasting and demonstrate that the meta-learning based approach outperforms individual forecasting models.

There are three main technical issues in the aforementioned approaches. First, most of the aforementioned approaches are based on so called Rice's structure introduced in [30], which represents only a particular case of meta-learning. The essential question of "how to define the LF model selection as a meta-learning problem?" is not well addressed. Second, significant ambiguity exists in metalearner selection. In Rice's structure, the key part of the model selection system is to let the metalearner establish an effective mapping from the task features to the optimal model recommendation. Existing approaches often select a classical classification algorithm to serve as the metalearner. Because different classifiers target different expertise areas, a weighting mechanism is needed to combine their assessments for making the final recommendation. Third, without a rigorous LF task set up criterion, the performance of the meta-learning based approach cannot be properly quantified. For example, only tens of toy cases are used in one study whereas hundreds of similar forecasting tasks are used in another. Consequently, the model selection accuracy can range anywhere from 20% to 90% depending on which test cases are used for quantifying the performance of a trained metalearner.

To overcome those technical issues, we propose a meta-learning based, automatic distribution system load forecasting model selection framework. The framework includes the following processes: feature extraction, candidate model preparation and labeling, offline training, and online model recommendation. Using user load forecasting needs and data characteristics as input features, multiple metalearners are used to rank the available load forecast models based on their forecasting accuracy. Then, a scoring-voting mechanism weights recommendations from each meta-leaner to make the final recommendations.

The contributions of this paper are considered threefold.

(1) First, we propose a generalized meta-learning framework with rigorous mathematical formulation for solving power system load forecasting model selection problems. The framework is automatic and extendable, and is also computationally efficient in online operation period.

(2) Second, based on the ensemble learning concept, we introduce a scoring-voting mechanism for combining the strength of multiple metalearners, which increases the model recommendation accuracy.

(3) Third, we develope a procedure to set up heterogeneous load forecasting tasks in distribution systems, and create a test case containing over 800 load forecasting tasks based on field data, which can be considered as a standard test case for follow-up studies.

## 2. Problem Formulation

The framework (see Fig. 1) consists of two layers: a base-learning layer and a meta-learning layer. In the base-learning layer, $J$ learning tasks are created. In the $J$ pairs of data samples, $\langle \mathbf{X}_j, \mathbf{y}_j \rangle$, $\mathbf{X}_j$ is the input time series data of an LF model with a dimension of $N_j \times M_j$, $\mathbf{y}_j$ is the actual load with a dimension of $N_j \times 1$, and $j \in [1, .., J]$. To conduct each LF task, we divide $\mathbf{X}_j$ into $\mathbf{X}_j^{train}$ and $\mathbf{X}_j^{test}$, and $\mathbf{y}_j$ into $\mathbf{y}_j^{train}$ and $\mathbf{y}_j^{test}$ so that $N_j - K_j$ samples are used for training and $K_j$ samples are used for testing.

For an LF task, there are $I_B$ LF models serving as candidate LF algorithms, as shown in the base-learning layer schematic in Fig. 1. Each LF model will be trained using $\langle \mathbf{X}_j^{train}, \mathbf{y}_j^{train} \rangle$ and tested using data set $\langle \mathbf{X}_j^{test}, \mathbf{y}_j^{test} \rangle$. The model accuracy is calculated using the root-mean-square error (RMSE) between $\hat{\mathbf{y}}_{j,i_B}^{test}$ and $\mathbf{y}_j^{test}$. The LF model with the smallest RMSE will be selected as the model to be used for this training task.

Thus, after we complete the training and testing for all $J$ LF tasks in the base-learning layer, the best performed LF model for each LF task, $\mathbf{\Phi}$, is considered known and labeled. The input feature matrix of the meta-learning layer, $\mathbf{F}$, has a dimension of $J \times D$. The input features of each LF task consist of two parts: input data statistics, $\mathbf{F}(j, 1: R)$ and technical requirements of the LF task $\mathbf{F}(j, R + 1: D)$.

As shown in the meta-learning layer schematic in Fig. 1, meta-data obtained from the base-learning layer, $\langle \mathbf{F}, \boldsymbol{\Phi} \rangle$, is divided into a training meta-data set $\langle \mathbf{F}^{train}, \boldsymbol{\Phi}^{train} \rangle$ and a testing meta-data set $\langle \mathbf{F}^{test}, \boldsymbol{\Phi}^{test} \rangle$. There are $I_M$ metalearners used, so $I_M$ sets of recommendations, $\langle \widehat{\boldsymbol{\Phi}}_{I_M} \rangle$, will be obtained. Then, a voting engine that weights $\langle \widehat{\boldsymbol{\Phi}}_{I_M} \rangle$ by predicted accuracy of each metalearner will be used to determine the optimal $\langle \widehat{\boldsymbol{\Phi}} \rangle$. In the following subsections, we will introduce the problem formulation of the base-learning layer LF model selection process and the meta-learning layer LF model recommendation mechanisms and illustrate the online application procedure.

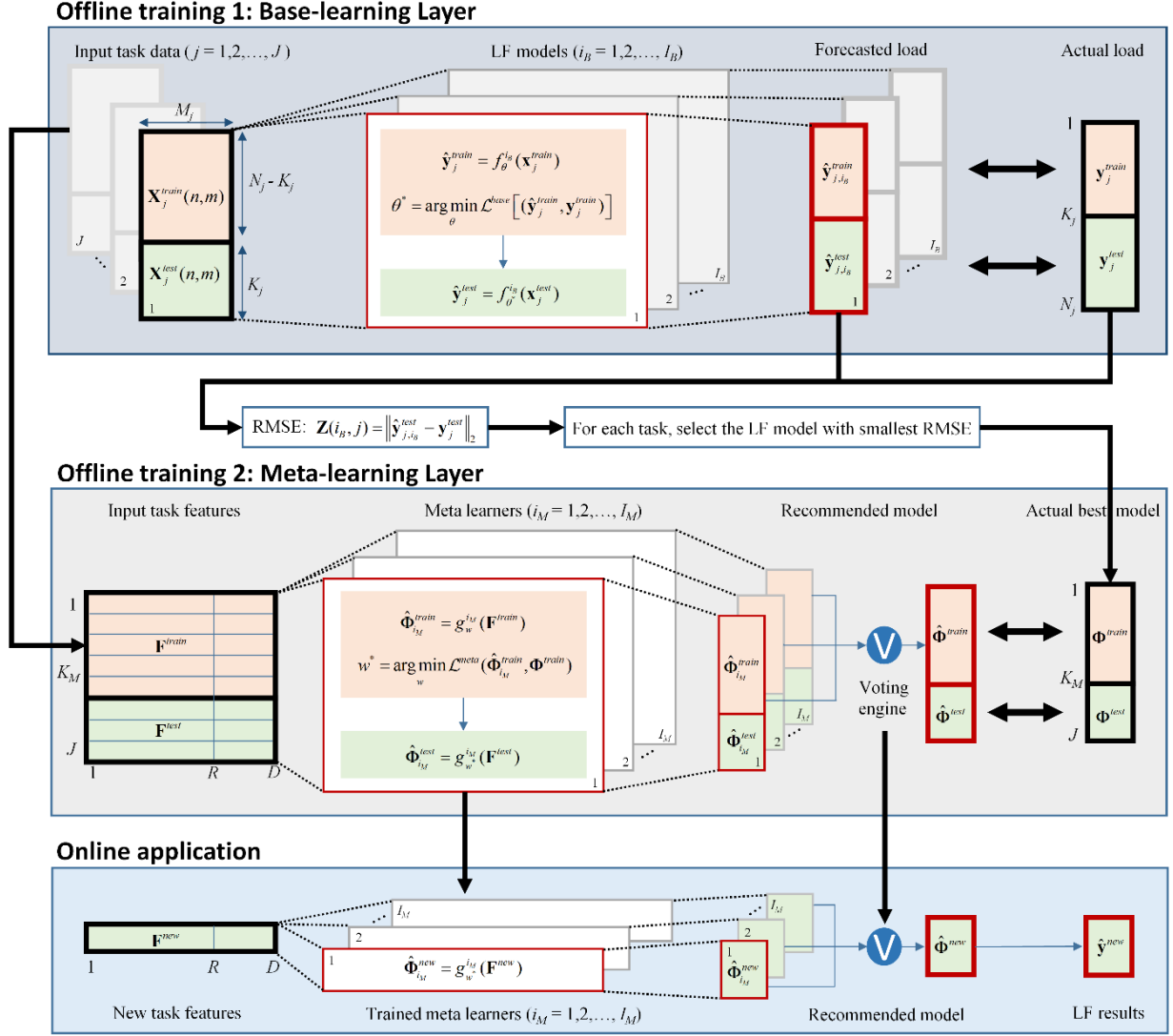

Fig. 1 Flowchart of the proposed meta-learning based LF model selection framework.

## 2.1 Base Learning Layer Problem Formulation

In the machine learning domain, power system LF problems belong to supervised machine learning. The set of training data $\mathbf{X}_j^{train}$ for the $j^{\text{th}}$ LF task can be represented as

$$\mathbf{X}_j^{train} = \begin{bmatrix} x_1^1 & x_1^2 & \cdots & x_1^M \\ x_2^1 & x_2^2 & \cdots & x_2^M \\ \vdots & \vdots & \ddots & \vdots \\ x_N^1 & x_N^2 & \cdots & x_N^M \end{bmatrix}_{(N_j - K_j) \times M} \tag{1}$$

where $x_n^m$ represents the $m^{\text{th}}$ attributes of the $n^{\text{th}}$ sample. Denote $\hat{y}_{j,i_B}^{train}$ as the forecasted load generated by LF model $i_B$ for the $j^{\text{th}}$ LF task, the base-learning layer problem can be formulated as

$$\hat{y}_{j,i_B}^{train} = f_\theta^{i_B}(\mathbf{X}_j^{train}) \tag{2}$$

$$\theta^* = arg \min_\theta \mathcal{L}^{base}(\hat{y}_{j,i_B}^{train}, \mathbf{y}_j^{train}) \tag{3}$$

where $\mathcal{L}^{base}$ is the loss function calculated as distance between the actual load $\mathbf{y}_j^{train}$ and the predicted load $\hat{y}_{j,i_B}^{train}$, and $\theta^*$ is the optimal parameters for LF model $i_B$. Once $\theta^*$ is obtained, the forecasting accuracy is measured by the RMSE errors on the testing data set, so we have

$$\hat{y}_{j,i_B}^{test} = f_{\theta^*}^{i_B}(\mathbf{X}_j^{test}) \tag{4}$$

$$\mathbf{Z}(i_B, j) = \left\| \hat{y}_{j,i_B}^{test} - \mathbf{y}_j^{test} \right\|_2 \tag{5}$$

The LF model with the highest accuracy among all $I_B$ LF models is selected as the recommended LF model for the $j^{\text{th}}$ LF task and its index is stored in $\mathbf{\Phi}(j)$.

## 2.2 Meta-learning Layer Problem Formulation

By summarizing cross-task knowledge into *meta-knowledge*, a metalearner can learn 'how to learn tasks' from known tasks in order to improve its performance in new tasks [31]. Meta-knowledge can be in different forms, for example, selecting algorithms or optimizers to solve different tasks [32] and finding initialization parameters for different machine learning models [18].

In this paper, meta-learning is used to find the best LF model for a LF task. The problem is formulated as

$$\hat{\mathbf{\Phi}}_{i_M}^{train} = g_w^{i_M}(\mathbf{F}^{train}) \tag{6}$$

$$w^* = arg \min_w \mathcal{L}^{meta}(\hat{\mathbf{\Phi}}_{i_M}^{train}, \mathbf{\Phi}^{train}) \tag{7}$$

where $g_w$ represents the *metalearner* with parameter $w$, $\mathcal{L}^{meta}$ is the loss function measuring the distance between the actual best LF model $\mathbf{\Phi}^{train}$ and the recommended LF model by meta-leaner $g_w^{i_M}$, $\hat{\mathbf{\Phi}}_{i_M}^{train}$.

Once the optimal parameters $w^*$ is determined, the performance of the metalearner will be tested on the testing set. The recommendation accuracy, $\eta_{i_M}$, is calculated as

$$\hat{\mathbf{\Phi}}_{i_M}^{test} = g_{w^*}^{i_M}(\mathbf{F}^{test}) \tag{8}$$

$$\eta_{i_M} = \frac{1}{J - K_M} \sum_{j=K_M+1}^{J} I_{[\hat{\mathbf{\Phi}}_{i_M}^{test}(j) = \mathbf{\Phi}^{test}(j)]} \tag{9}$$

Because multiple metalearners are used to cover the diversity in LF tasks, recommendations from different metalearners, $\hat{\mathbf{\Phi}}_{i_M}, i_M \in [1, ..., I_M]$, need to be weighted through a *scoring-voting* mechanism in order to obtain the final model recommendation $\hat{\mathbf{\Phi}}$.

The accuracy of the final model recommendation, $\eta$, is calculated as

$$\eta = \frac{1}{J - K_M} \sum_{j=K_M+1}^{J} I_{[\hat{\mathbf{\Phi}}^{test} = \mathbf{\Phi}^{test}(j)]} \tag{10}$$

## 2.3 Online Application and Framework Extension

After the training is finished, the framework can be applied online for recommending one or a few LF models for new LF tasks. First, the feature set of the new LF task, $\mathbf{F}_{new}$, is calculated. Then, recommendations from all metalearners, $g_{w^*}^{i_M}(\mathbf{F}_{new}), i_M \in [1, ..., I_M]$, will be sent to the voting engine to obtain the final LF model recommendation, $\hat{\mathbf{\Phi}}^{new}$. The online application involves only forward calculation so it is very computationally efficient. The main advantage of the meta-learning based approach is its extendibility because a user can readily incorporate new task samples, LF models, meta-features and metalearners into the existing framework, making it scalable and low maintenance

## 3. Implementation Setup

This section introduces the implementation setup of the proposed meta-learning LF model selection framework.

### 3.1 LF Task setup

To learn how to select the best LF models for unseen LF tasks, it is critical for the metalearners to be trained and tested on a large amount of heterogeneous distribution system LF tasks. In this paper, we consider that LF Tasks differ from one another in five aspects: data granularity, data length, forecasting horizon, exogenous factors, and load aggregation level. Thus, a building-level day-ahead LF task with 1-year hourly load and temperature data sets as inputs can be described by a red dashed line in the 5-dimensional radar chart in Fig. 2. By randomly select values of the variable representing the five LF task features, a wide range of heterogeneous LF tasks can be created.

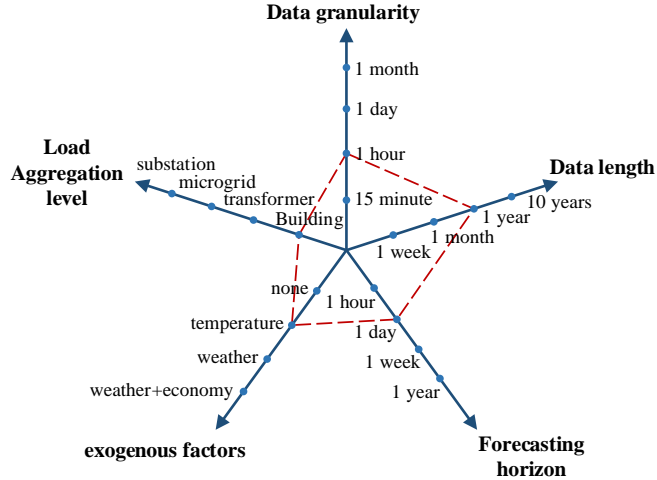

Fig. 2 Five main features representing heterogeneous LF tasks

### 3.2 Selection of Candidate LF Models

Many LF models have been developed in literature for solving different LF tasks. In this paper, four LF models commonly used for forecasting distribution system loads are selected: Seasonal Autoregressive Integrated Moving Average (SARIMA), Long Short-Term Memory (LSTM), Support Vector Regression (SVR) and Similar Day (SD). Note that for a given LF model, one can select different model structures in order to achieve the best performance in a given LF task. Therefore, when preparing the LF candidate models, we consider 6 SARIMA and 2 LSTM model structures to demonstrate that the proposed meta-learning framework is also effective in selecting model structures. The 10 candidate LF models are shown in Table I.

- *SARIMA* models time series with seasonal characteristics [33][34]. The basic structure of SARIMA is the same as ARIMA model

$$y_t = \varphi_1 y_{t-1} + \varphi_2 y_{t-2} + \cdots + \varphi_p y_{t-p} - \theta_1 \varepsilon_{t-1} - \theta_2 \varepsilon_{t-2} - \cdots - \theta_q \varepsilon_{t-q} \tag{11}$$

where $p$, $q$ determines the structure of the model, and $\varphi$, $\theta$ are the coefficients. Here we use 6 different structures of SARIMA model, shown in Table I. For example, SARIMA (2,1,1)(2,1,1) refers to The SARIMA model with $p$=2, $q$=1, and 1-order difference for both the trend component and the seasonal component. Please note that since in each SARIMA model we are using the same $p$ and $q$ values for both the trend component and the seasonal component, hereinafter we will use SARIMA(2,1,2) to represent SARIMA(2,1,2)(2,1,2) for simplicity.

- *LSTM* is an upgraded version of Recurrent Neural Network equipped with long-term memory capability [35]. The key structural parameter for LSTM is the number of hidden units. Here we consider 2 typical structures: 125 and 200 hidden units of LSTM.

- *SVR* is a classical regression method by finding a hyperplane to separate high-dimension data [36]. Here we introduce SVR to mainly solve forecasting tasks with exogenous factors. The kernel function we use for SVR is Gaussian kernel.

- *SD* tries to find the most similar day in the historical data pool for the forecasting day, considering calendar information and exogenous factors [37]. SD can be used when historical data is not sufficient for training complex forecasting models. Let $\Delta T = T_{fore} - T_{hist}$, where $T_{fore}$, $T_{hist}$ represent the time indexes of the forecasting day and the historical day. Then the similarity between the forecasting day and the historical day, $\gamma$, is calculated as

$$\gamma = \frac{\beta_1^{(1-C)\text{mod}(\Delta T/7)}\beta_2^{(1-C)\text{floor}(\Delta T/7)}\beta_3^{(1-C)\text{floor}(\Delta T/365)}}{\left\| \mathbf{x}_j(T_{fore},:) - \mathbf{x}_j(T_{hist},:) \right\|_2} \tag{12}$$

where $\beta_1, \beta_2, \beta_3 \in (0,1)$, $C$ is binary variable that equal 1 when mod(t/365)=0 and equal to 0 otherwise. In (12), the numerator measures the calendar similarity and the denominator quantifies the distance of the exogenous factors, so the historical day with the largest $\gamma$ will be selected as the SD for the forecasting day.

Table I Candidate LF model summary

| Number | Candidate LF models | Number | Candidate LF models |
|---|---|---|---|
| 1 | SARIMA (2,1,1)(2,1,1) | 6 | SARIMA (5,1,5)(5,1,5) |
| 2 | SARIMA (3,1,3)(3,1,3) | 7 | LSTM(125) |
| 3 | SARIMA (4,1,2)(4,1,2) | 8 | LSTM(200) |
| 4 | SARIMA (4,1,4)(4,1,4) | 9 | SVR |
| 5 | SARIMA (5,1,2)(5,1,2) | 10 | SD |

### 3.3 Candidate Model Labeling

To find the best performed LF model $\mathbf{\Phi}(j)$ for LF task $j$, (2)-(5) are repeated $L_j$ times with different training and testing data splits. This allows us to obtain an estimation of the distribution of the top 1 LF model, $\mathbf{\Omega}_j$. One can iterate the process until the distribution is stabilized. Pearson correlation coefficient [38], $P_{cc}$, is used as the stopping criterion. $P_{cc}$ is a statistic that measures the correlation between two vectors. The iteration will stop when $P_{cc}$ between $\mathbf{\Omega}_j(L_j)$ and $\mathbf{\Omega}_j(L_j - 10)$ is larger than 0.95. Note that this step is critical for removing the uncertainty in selecting the best LF model for each LF task. The pseudocode of determining $\mathbf{\Phi}(j)$ is shown in Algorithm 1.

---

**Algorithm 1:** Candidate model labeling for LF task $j$

---

    **Input:** $\mathbf{X}_j$
    **Output:** $\mathbf{\Phi}(j)$
    Initialize $P_{cc} = 0$, $L_j = 1$
    **while** ($P_{cc} <$0.95) **do**
        increase $L_j$ by 10
        **for** 1,2,…, $L_j$ **do**
            Randomly split $\mathbf{X}_j$ into $\mathbf{X}_j^{train}$ and $\mathbf{X}_j^{test}$
            Train each candidate LF model based on (2)(3)
            Test each candidate LF model based on (4)(5)
            Label the LF model with the smallest RMSE as top 1
        **end**
        Calculate $\mathbf{\Omega}_j(L_j)$
        Calculate $P_{cc}$ between $\mathbf{\Omega}_j(L_j)$ and $\mathbf{\Omega}_j(L_j - 10)$
    **end**
    **return** $\mathbf{\Phi}(j) =$ LF model with the highest frequency in $\mathbf{\Omega}_j(L_j)$

---

## 3.4 Meta features of LF tasks

In this paper, a feature set $\mathbf{F}$ containing 16 features (See Table II) is used to characterize each task. Feature 1-6 describe the basic features of a LF task. Feature 7-16 are statistics characterizing the historical load profile.

Table II Features to describe LF tasks

| Number | Candidate LF models | Number | Candidate LF models |
|--------|---------------------|--------|---------------------|
| 1 | Data length | 9 | Minimum |
| 2 | Number of weather features | 10 | Standard deviation |
| 3 | Data granularity | 11 | Kurtosis |
| 4 | Forecasting horizon | 12 | Skewness |
| 5 | Number of customers | 13 | Fickleness |
| 6 | Load type | 14 | H-ACF |
| 7 | Mean | 15 | H-PACF |
| 8 | Maximum | 16 | Periodicity |

*Kurtosis* and *Skewness* can be calculated by (13) and (14), where $\sigma, \bar{y}$ are the standard deviation and mean value of the historical load profile.

$$Kurtosis = \frac{1}{N\sigma^2} \sum_{n=1}^{N} [\mathbf{y}(n) - \bar{y}]^4 \tag{13}$$

$$Skewness = \frac{1}{N\sigma^2} \sum_{n=1}^{N} [\mathbf{y}(n) - \bar{y}]^3 \tag{14}$$

*Fickleness* measures the ratio of a time series crossing its mean value and is calculated as

$$Fickleness = \frac{1}{N} \sum_{n=2}^{N} I_{\{sign[\mathbf{y}(n-1)-\bar{y}]=sign[\mathbf{y}(n-1)-\bar{y}]\}} \tag{15}$$

*Highest Autocorrelation Function* (H-ACF) and *Highest Partial Autocorrelation function* (H-PACF) measure the self-correlation features of the load profile, which is especially useful for determining the structure of a SARIMA model. *Periodicity* of the load profile is usually related to the data granularity. For example, periodicity is usually 24 or 168 for an hourly load profile and 30 for a daily load profile.

### 3.5 Metalearner selection and scoring-voting mechanism

A metalearner maps the meta-task features $\mathbf{F}$ to the best LF models $\mathbf{\Phi}$ for a given LF task. This makes it essentially a classification problem. Thus, in this paper, 4 classical classification algorithms with different strength are selected: *Random Forest* (RF) [39], *K-Nearest Neighbor* (KNN) [40], *Naïve Bayesian* (NB) [41] and *Linear Discrimination* (LD) [42].

To combine the four recommendations from the four metalearners into one, an *scoring-voting* mechanism based on ensemble-learning concept is developed, as shown in Fig. 3.

Note that each classifier accomplishes its classification based on an internal scoring procedure. For example, NB calculates the posterior probability of each class as their scores, while RF counts the voting results from its wrapped decision trees. The metalearner $i_M$ selects the candidate model with the highest score $S_{i_M}$ as its output. A higher score means a stronger belief of the classifier on its output, therefore leading to a higher classification accuracy.

We then establish the relationship $h_{i_M}$ between the score $S_{i_M}$ and the classification accuracy $\eta_{i_M}$ for each metalearner, based on their performance on the testing LF tasks

$$\eta_{i_M} = h_{i_M}(S_{i_M}) \tag{16}$$

When dealing with a new task online, we first transfer the scores $S_{i_M}$ of each metalearner to their accuracy level $\eta_{i_M}$ using (16), and then select the candidate LF model with the highest accuracy level as the final choice $\hat{\mathbf{\Phi}}^{new}$.

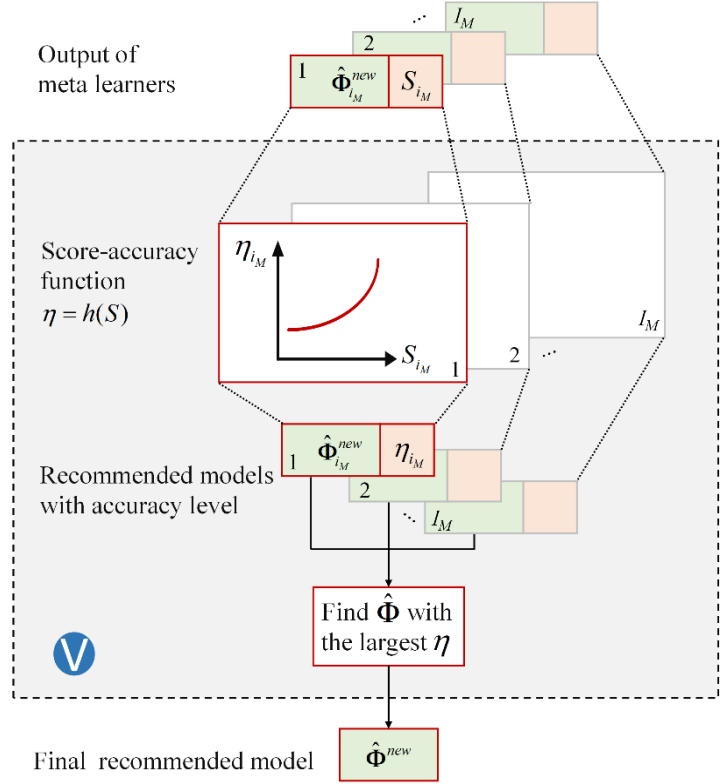

Fig. 3 Scoring-voting system to combine different metalearners.

## 4. Experiment Setup and Results

This section presents experiments and results for evaluating the performance of the meta-learning based distribution LF model selection framework.

### 4.1 LF Task Setup

Creating a large amount of heterogeneous LF tasks is crucial for training the metalearner. As summarized in Table III, we consider five key forecasting requirements: the aggregation level, the number of weather features, the historical data lengths, forecasting horizon, and granularity of data, each of which represents one of the five dimensions shown in Fig. 3.

Table III Heterogeneous Forecasting Task Selection

| LF Tasks | Aggregation Level | Weather Features | Data Length | Forecasting Horizon | LF Data Granularity |
|---|---|---|---|---|---|
| Building-level | 1 residential/ 1 commercial | 0 1 | 1 month 6 months 1 year | 4h 24h 168h | 15 min 30 min 1 h |
| Distribution Transformer | 3-10 residential/ 2-4 commercial | | | | |
| Community Microgrid | 50-300 users | | | | |
| Distribution Feeder | 1000-2000 users | 0, 1, 12 | | 4h, 24h, 168h 30 days | 1h 1 day |

Residential and commercial load profiles are from two data sources: 15-minute and 30-minute smart meter data sets collected from utilities in North Carolina areas and 1-minute data sets from Pecan Street data repository [43]. Hourly weather data is downloaded from the National Oceanic and Atmospheric Administration (NOAA) website [44]. In this paper, 846 LF tasks are constructed by the exhaustive combination of the LF requirements within the ranges given in Table III and by following the following additional considerations:

- From the building-level to the feeder-level, LF tasks are designed for two main classes of loads: residential and commercial. Industrial load and agriculture loads are not considered.
- From the building-level to the microgrid-level, we focus on short-term LF because the goal of such LF tasks is usually to support demand response management programs in operation [45].
- At the feeder-level, mid-term LF tasks are also considered. Also, we assume that we have up to 12 available weather features from weather service providers.
- In practice, historical data available for a distribution LF task can be very short so we considered three cases to cover the data availability issue: 1-month, 6-month and 1-year. Also, because weather data may not always be available in a distribution LF task, we consider the case with zero weather feature.

### 4.2 LF Model Selection in the Base-learning Layer (all task)

Following the LF model selection process introduced in Section III, the statistically best performed LF models for 846 LF tasks are selected. In Fig. 4(a), we use the result from one of the 846 tasks as an example to illustrate the best LF model selection process. When the number of iteration increases, the distribution of the top 1 LF model starts to stabilize. After 60 iterations, the Pearson coefficient is above 0.98, so the iteration stops and model 9 is selected as the best performed model for this task. The box plot in Fig.4(b) shows that approximately 50% of the 846 LF tasks require 20 to 40 iterations to identify the best LF model using the proposed method.

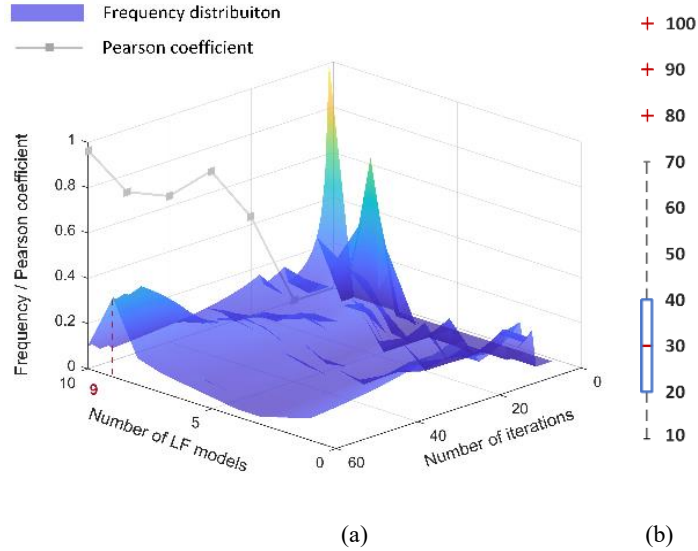

(a)                                                                        (b)

Fig. 4. (a) An example of the LF model selection process in the base-learning layer. (b) boxplot of required iterations to label each of the 846 LF tasks.

The model selection results of all the 846 LF tasks are summarized in Fig. 5. Note that if the historical data of a LF task is insufficient to train a LF model, we consider the LF model as an infeasible model for accomplishing this LF task.

The results show that model 7, LSTM(125), is the most frequently-selected model (173 out of 846 tasks) and model 10, the SD model, has by far the shortest training time and lowest data requirement among all options, but the mean and variance of its forecasting error are larger than other LF models.

Because the SARIMA-based approach requires significant amount of historical data to train, SARIMA-based LF models have higher failure counts. Among the six SARIMA models selected, SARIMA(5,1,5) (model 6) failed in 64 tasks and SARIMA(4,1,4) (model 4) failed in 40 tasks. However, on average, they tend to have a higher forecasting accuracy for the tasks with sufficient training data sets.

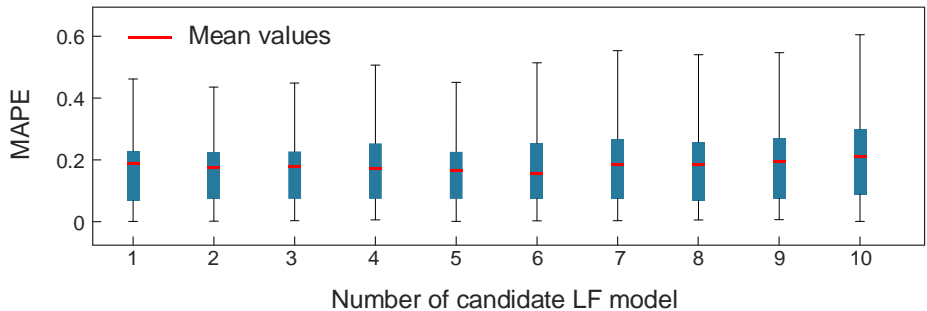

| LF model | #1 | #2 | #3 | #4 | #5 | #6 | #7 | #8 | #9 | #10 |
|---|---|---|---|---|---|---|---|---|---|---|
| **Top 1 count** | 120 | 28 | 17 | 30 | 15 | 58 | **173** | 124 | 130 | 151 |
| **Failure count** | **0** | 17 | 8 | 40 | 15 | 64 | **0** | **0** | **0** | **0** |
| **Time cost (s)** | 2.7 | 4.5 | 4.4 | 5.7 | 4.8 | 6.7 | 34.2 | 40.6 | 0.9 | **0.04** |
| **Mean MAPE** | 0.193 | 0.189 | 0.191 | 0.190 | 0.189 | **0.188** | 0.204 | 0.219 | 0.219 | 0.222 |
| **SER** | 1.49 | 1.46 | 1.50 | 1.44 | **1.40** | 1.42 | 1.56 | 1.44 | 1.62 | 1.79 |

Fig. 5. LF model selection results on all 846 tasks.

To further quantify the distance between different LF models, we define the *System Error Ratio* (SER) as

$$SER = \frac{E_{select}}{E_{best}} \tag{17}$$

where $E_{select}$ is the forecasting RMSE of a selected candidate model on a specific task, and $E_{best}$ is the forecasting RMSE of the actual best model among the candidates on this task. SER measures the distance between the selected model and the actual best model for each LF task.

Figure 6 shows the SERs of different ranking candidates on all the 846 LF tasks. We can see that the performance of top 2-4 models on most LF tasks are very close to the best model identified (i.e., SERs is close to 1). However, the tasks outside the top 5 can perform poorly or even fail, leading to a large SER value. Clearly, the performance of different LF models can vary significantly when performing a task. This demonstrates the importance of the LF model selection process.

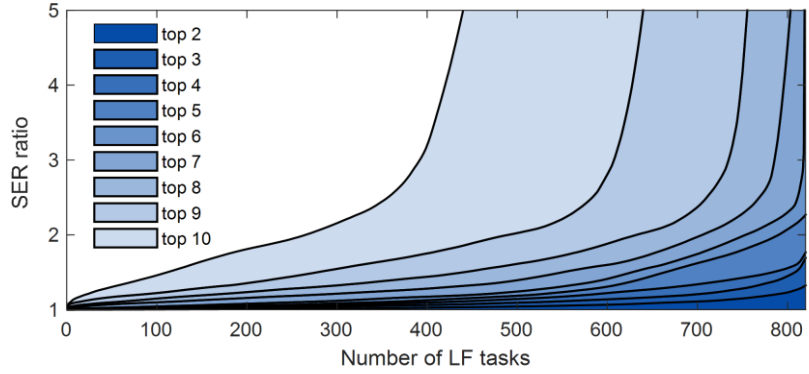

Fig. 6. SER values of candidate LF models under different rankings.

### 4.3 LF Tasks Similarity Evaluation

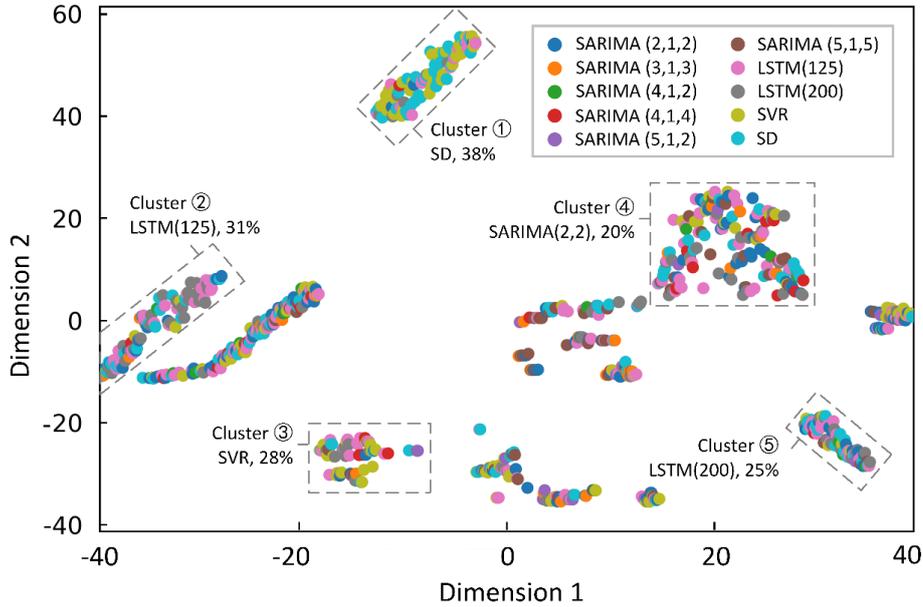

Fig. 7. t-SNE visualization of LF tasks labeled by their best performance models

Recall that at the meta-learning level, the input feature matrix of the meta-learning layer, **F**, consists of two parts: input data statistics and technical requirements of the LF task. We apply T-distributed Stochastic Neighbor Embedding (t-SNE) [46] to visualize the similarity among the LF tasks. Through nonlinear dimension reduction, the originally 16-dimensional **F** is reduced to a two-dimension matrix so distancing-based clustering method can be used to identify the five representative clusters, as shown in Fig. 7 and Table IV.

Table IV Values of the main task features of each cluster

| Cluster | Load Level | Weather Feature | Historical Data Length (day) | Forecasting Horizon (h) | Data Granularity (h) |
|---------|-----------|-----------------|------------------------------|-------------------------|----------------------|
| ① | 1000-2000 | 0,1,12 | 30,180,360 | 720 | 24 |
| ② | 1000-2000 | 0,1,12 | 30 | 4-168 | 1 |
| ③ | 1-10 | 0,1 | 30 | 4-24 | 0.25-1 |
| ④ | 50-300 | 0,1 | 30,180,360 | 4-168 | 0.25-1 |
| ⑤ | 1500-2000 | 0,1,12 | 180,360 | 4-168 | 1 |

In Fig. 7, each colorized dot represents a LF task labeled by its best performance model. After applying t-SNE visualization, similar LF tasks are more likely to appear near each other whereas dissimilar LF tasks appear far apart with each other.

The results show that clusters 1, 2, and 5 represent feeder-level LF tasks, among which cluster 1 represents mid-term forecasting with daily data granularity with SD as the dominant model. This is because although historical data is insufficient to train a complicate model in those cases, the load profile is normally stable and exhibits clear periodicity. Clusters 2 and 5 represent short-term forecasting with weather features available, making LSTM the best model in most cases. Cluster 3 represents short-term LF tasks at the building- and transformer- levels, where the load profiles are highly volatile, making SVR more frequently picked. Cluster 4 represents short-term microgrid-level LF tasks with little weather information, making LF tasks more often formulated as time series analysis problems suitable for SARIMA.

### 4.4 Metalearner Training, Validation, and Testing Results

To train, validate, and test the LF model selection metalearner, we randomly split 846 tasks into three groups: *training* (70%), *validation* (20%), and *testing* (10%), respectively. The program, developed using Matlab 2020a, is executed on an Intel i9-9900K, 64GB desktop. The meta-learner training takes approximately 5.3s. After the four metalearners are trained on the training set, their performance will be validated on the validation set. Then, LF tasks will be randomly selected from the testing set to evaluate the performance for a metalearner to provide online LF model recommendation (see next subsection E).

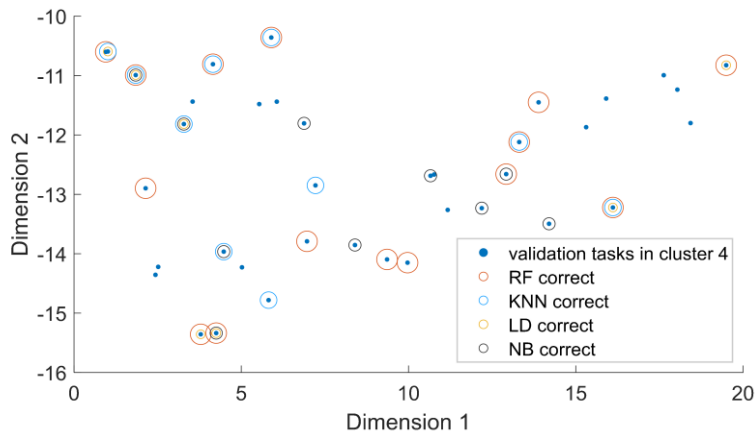

Fig. 8. Testing results of trained metalearners in cluster ④.

Figure 8 shows the validation results on LF tasks in cluster 4 as an example. Each blue dot represents a validation LF task in cluster 4. If a metalearner successfully identifies the best LF model for a LF task, a circle of its specified color will be placed around the dot. Thus, "no circle around a dot" means that none of the four metalearners identified the best model; "multiple circles around a dot" means that more than one metalearners have successfully identified the best LF model.

As discussed in Section IIX, to value the strength of each metalearner, a scoring-voting mechanism is developed to weight recommendations from each metalearner. As shown in Fig. 9, in general, the score $S$ is proportional to the metalearner classification accuracy calculated by (9). This allows the recommendation with the highest score to be used as the final recommended model. As shown in Table V, the classification accuracy of the proposed scoring-voting mechanism is 46%, which is 36% higher than the baseline (random selection) and 8-13% higher than that of an individual metalearner. Finally, as shown in Table VI, the proposed scoring-voting meta-learning mechanism can significantly reduce the forecasting error compared to any single LF model as well as successfully avoid selecting LF models that cannot perform the LF task.

In additional to recommending the best LF model for a LF task, the proposed meta-learning framework can also rank all candidate models so that the second-best or third-best LF models can be recommended. As shown in Table VII, the average SER of the top three models are all lower than the single models listed in Fig. 5, whereas the classification accuracy of the three models are all above the baseline 10% (a sum of accuracy is 76%) with little or no failure. This means the meta-learning system can recommend on average three high-quality LF models for each LF task.

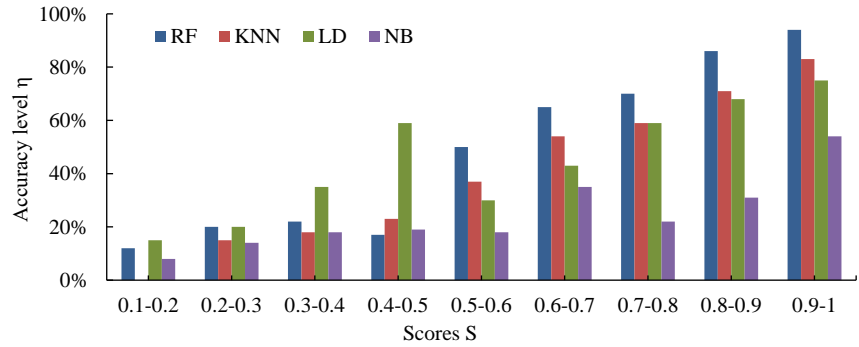

Fig. 9.   Metalearner accuracy versus Score  $S$.

Table V Metalearner accuracy comparison

|  | RF | KNN | LD | NB | Scoring-voting | Baseline |
|---|---|---|---|---|---|---|
| Accuracy | 38% | 35% | 33% | 33% | **46%** | 10% |

Table VI Performance comparison on SER ratio and MAPE

|  | Average SER | Average MAPE | Failure Count |
|---|---|---|---|
| Proposed meta-learning mechanism | **1.14** | **0.143** | 0 |
| Best-performed single LF model | 1.40 | 0.188 | 0 |

Table VII Performance of LF models on different rankings

| Ranking | 1 | 2 | 3 | 4 | 5 | 6 | 7 | 8 | 9 | 10 |
|---|---|---|---|---|---|---|---|---|---|---|
| **Classification accuracy** | 46% | 17% | 13% | 6% | 4% | 3% | 3% | 3% | 2% | 3% |
| **SER** | 1.14 | 1.27 | 1.34 | 1.46 | 4.18 | 2.89 | 4.48 | 3.61 | 2.61 | 3.09 |
| **Failure count** | 0 | 0 | 2 | 10 | 10 | 12 | 12 | 17 | 14 | 11 |

To analyze the robustness of the proposed method, we calculate the SER ratio of the top-1 recommendations on the 40 failure tasks (24% of all cases) where the proposed method fails to hit the actual best model among its top 3 recommendations. As shown in Fig. 10, the average SER value is 1.19 with no infeasible models. This shows that, although in some cases the proposed method fails to identify the actual best model, the top-1 model recommendation is still effective with performance close to the actual best model.

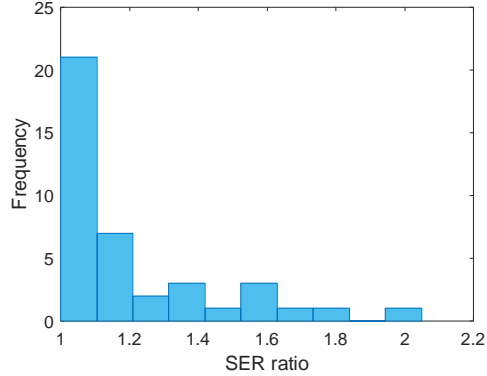

Fig. 10. Histogram of the SER ratio of the top-1 recommendations on the 40 failure tasks.

### 4.5 Online Testing Performance

Two LF tasks, with features presented in Table VIII, are used to illustrate the online operation procedure. Task 1 is a transformer-level short-term (24 hours ahead) LF task with 30 days, 15-minute historical load data; Task 2 is a feeder-level short-term (1-week ahead) LF task using 6-month, 1-hour historical data. The top one model will be recommended in task 1 and the top three model will be recommended in task 2.

Table VIII Performance of LF models on different rankings

|  | Task 1 | Task 2 |
| --- | --- | --- |
| **Data granularity (hour)** | 0.25 | 1 |
| **Historical data length (day)** | 30 | 180 |
| **Number of weather factors** | 0 | 0 |
| **Forecasting horizon (hour)** | 24 | 168 |
| **Load level (# of user)** | 5 | 1100 |

Figure 11 uses results of task 1 as an example of using the scoring-voting mechanism to identify the top-1 model. Firstly, calculate 16 meta features of task 1 and input them to the trained meta-learners. Then, the meta-learners compute a score for each candidate LF model for quantifying its feasibility when solving task 1. The models with the highest scores are selected. The scores will then be converted to the accuracy level (see Fig. 9) and the LF model with the highest accuracy level is marked as the top-1 choice.

Results are shown in Fig. 12, Tables IX and X. We can see that for task 1 the system successfully recommends the actual best model SD. For task 2, three SARIMA models with similar performance are recommended as the top 3, with the actual best model SARIMA(5,1,5) ranked as the second.

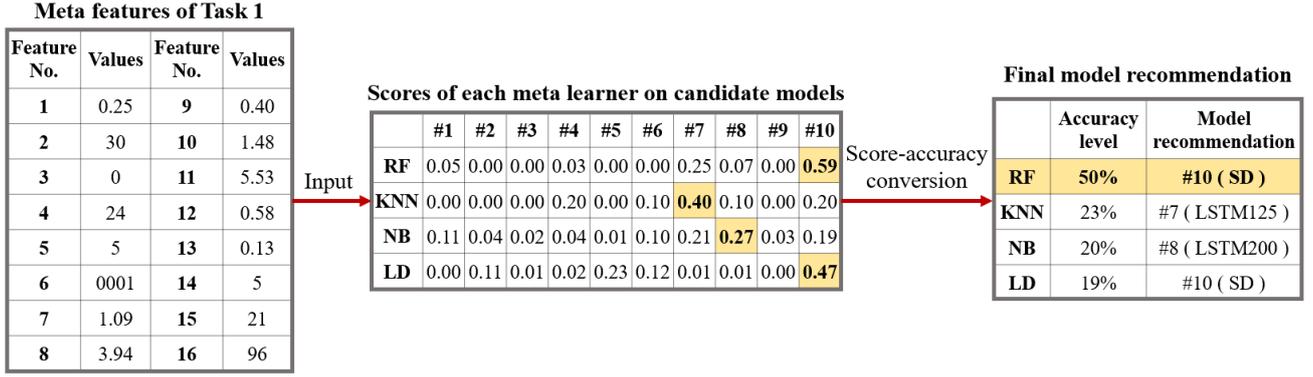

Fig. 11. An example of the scoring-voting process based on task 1 results.

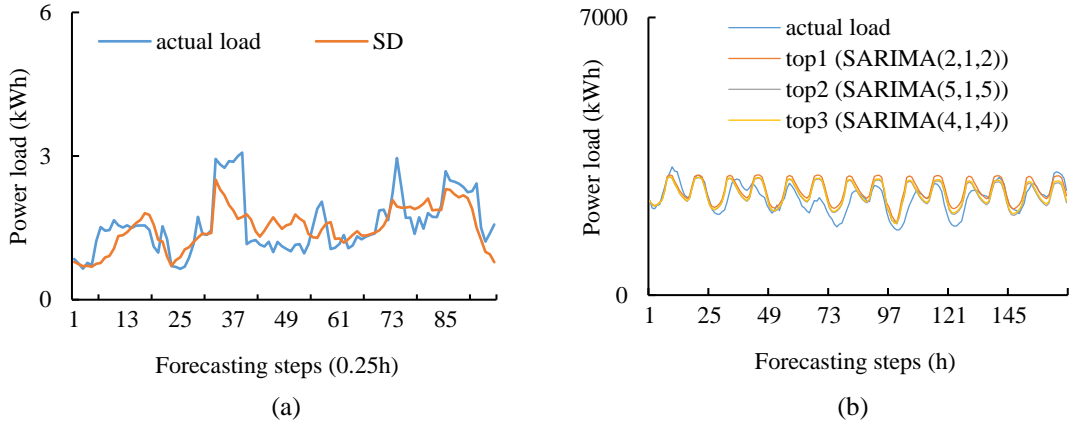

(a)                                                         (b)

Fig. 12. (a) LF results of Task 1, (b) LF results of Task 2.

Table IX forecasting RMSE and MAPE of the 10 candidate models

| LF model | | #1 | #2 | #3 | #4 | #5 | #6 | #7 | #8 | #9 | #10 |
|---|---|---|---|---|---|---|---|---|---|---|---|
| Task 1 | RMSE | 0.93 | 0.90 | 0.89 | 0.89 | 0.87 | 0.75 | 0.69 | 0.85 | 1.57 | 0.39 |
| | MAPE | 44% | 42% | 41% | 41% | 40% | 32% | 27% | 58% | 67% | 19% |
| Task 2 | RMSE | 244 | 301 | 400 | 207 | 210 | 287 | 844 | 1509 | 453 | 674 |
| | MAPE | 8% | 11% | 14% | 6% | 6% | 10% | 31% | 53% | 15% | 24% |

Table X Recommendations and forecasting accuracy in online application

| | Task 1 | Task 2 |
|---|---|---|
| Actual best model | SD | SARIMA(5,1,5) |
| Top-1 Recommendation & MAPE | SD, 19% | SARIMA(2,1,1), 8% |
| Top-2 Recommendation & MAPE | / | SARIMA(5,1,5), 6% |
| Top-3 Recommendation & MAPE | / | SARIMA(4,1,4), 6% |

Note that the most time-consuming part of the proposed meta learning system is labeling candidate models, where all candidate models are executed for completing all the sample tasks exhaustively. However, this part can be done once for all during offline training. The online procedure is simply a numerical calculation of task features and a forward application of the trained metalearners, which takes only about 0.2 s to complete using a desktop computer. Therefore, this approach are suitable for real-time applications. Users can specify a threshold for triggering system updates to add new forecasting tasks to the training set and retrain the metalearner. The main advantage of applying such a highly extendable system is that users can always add new features and candidate LF models to the meta-learning framework and avoid updating the whole system from scratch.

## 5. Conclusions

In this paper, we presented a meta-learning based LF model selection framework for handling heterogeneous forecasting tasks in distribution networks. Each metalearner will learn to select the LF model with the best performance for a given LF task in the offline training. The score-voting mechanism will learn to weight recommendations from different metalearners based on their strength in identifying the top candidate models. The resultant system recommends on average up to three effective LF models for each given LF task. Simulation results show that the top one recommendation has 46% chance and the top three recommendations have 76% chance to identify the actual best LF model. Compared with the best-performed single LF model, the proposed model selection framework can reduce the average MAPE from 0.188 to 0.143, and SER ratio from 1.40 to 1.14. The mechanism is highly scalability and extendibility because it allows users to introduce new features or candidate models.

Our future work will focus on the feature engineering to further improve the model selection accuracy. Also, more complex LF tasks considering the influences of renewable energy penetration and demand response will be created and incorporated into the tese cases.

## 6. Acknoledgement

This research is supported by the U.S. Department of Energy's Office of Energy Efficiency and Renewable Energy (EERE) under the Solar Energy Technologies Office Award Number DE-EE0008770

## 7. References

1. Tao Hong, Pierre Pinson, Yi Wang, Rafal Weron, Dazhi Yang, and Hamidreza Zareipour, "Energy Forecasting: A Review and Outlook," IEEE Open Access Journal of Power and Energy, 2020, 7:376-388.
2. Yi Wang, Gabriela Hug, Zijie Liu, and Ning Zhang, "Modeling Load Forecast Uncertainty Using Generative Adversarial Networks," Electric Power Systems Research, 2020, 189:106732.
3. Hafeez, Ghulam, Khurram Saleem Alimgeer, and Imran Khan. "Electric load forecasting based on deep learning and optimized by heuristic algorithm in smart grid." Applied Energy 269 (2020): 114915.
4. Cai, Mengmeng, Manisa Pipattanasomporn, and Saifur Rahman. "Day-ahead building-level load forecasts using deep learning vs. traditional time-series techniques." Applied Energy 236 (2019): 1078-1088.
5. Yi Wang, Ning Zhang, Yushi Tan, Tao Hong, Daniel Kirschen, and Chongqing Kang, "Combining Probabilistic Load Forecasts," IEEE Transactions on Smart Grid, 2019, 10(4):3664-3674.
6. Yi Wang, Dahua Gan, Mingyang Sun, Ning Zhang, and Chongqing Kang, "Probabilistic Individual Load Forecasting Using Pinball Loss Guided LSTM," Applied Energy, 2019, 235: 10-20.
7. Shu Zhang, Yi Wang, Yutian Zhang, Dan Wang, and Ning Zhang, "Load Probability Density Forecasting by Transforming and Combining Quantile Forecasts," Applied Energy, 2020, 277:115600.
8. Wu, Zhuochun, et al. "A hybrid model based on modified multi-objective cuckoo search algorithm for short-term load forecasting." Applied energy 237 (2019): 896-909.
9. Zhang, Jinliang, et al. "Short term electricity load forecasting using a hybrid model." Energy 158 (2018): 774-781.
10. Yi Wang, Qixin Chen, Mingyang Sun, Chongqing Kang and Qing Xia, "An Ensemble Forecasting Method for the Aggregated Load with Subprofiles," IEEE Transactions on Smart Grid, 2018, 9(4): 3906-3908.
11. Von Krannichfeldt, Leandro, Yi Wang, and Gabriela Hug. "Online Ensemble Learning for Load Forecasting." IEEE Transactions on Power Systems (2020).
12. Hu, Yi, et al. "Short-term load forecasting using multimodal evolutionary algorithm and random vector functional link

network based ensemble learning." Applied Energy 285 (2021): 116415.

13. M. S. Kandil, S. M. El-Debeiky, and N. E. Hasanien, "Long-term load forecasting for fast developing utility using a knowledge-based expert system," *IEEE Trans. on Power Syst.*, vol. 17, no. 2, Aug. 2002.

14. M. S. Kandil, S. M. El-Debeiky, and N. E. Hasanien, "The implementation of long-term forecasting strategies using a knowledge-based expert system: part-II," *Electric Power Systems Research*, vol. 58, no. 1, pp. 19-25, May 2001.

15. S. M. R. Kazemi, M. M. Seied Hoseini, S. Abbasian-Naghneh, and S. H. A. Rahmati, "An evolutionary-based adaptive neuro-fuzzy inference system for intelligent short-term load forecasting," *International transactions in operational research*, vol. 21, no. 2, Mar. 2014.

16. S. H. Liao, "Expert system methodologies and applications—a decade review from 1995 to 2004," *Expert systems with applications*, vol. 28, no. 1, pp. 93-103, Jan. 2005.

17. S. Thrun, L. Pratt, "Learning to learn: Introduction and overview," in *Learning to learn*, Boston, MA. Springer, 1998, pp. 3-17.

18. C. Finn, P. Abbeel, and S. Levine, "Model-agnostic meta-learning for fast adaptation of deep networks," *arXiv preprint*, arXiv:1703.03400, 2017.

19. C. Cui, T. Wu, M. Hu, J. D. Weir, and X. Li, "Short-term building energy model recommendation system: A meta-learning approach," *Applied energy*, vol. 172, pp. 251-263, Jun. 2016.

20. M. Feurer, J. T. Springenberg, and F. Hutter, "Initializing bayesian hyperparameter optimization via meta-learning," in *Twenty-Ninth AAAI Conference on Artificial Intelligence*, Feb. 2015.

21. C. Lemke, M. Budka, and B. Gabrys, "Metalearning: a survey of trends and technologies," *Artificial intelligence review*, vol. 44, no. 1, pp. 117-130, Jun. 2015.

22. Cui, Can, et al. "Short-term building energy model recommendation system: A meta-learning approach." Applied energy 172 (2016): 251-263.

23. Li, Wenqiang, et al. "Meta-learning strategy based on user preferences and a machine recommendation system for real-time cooling load and COP forecasting." Applied Energy 270 (2020): 115144.

24. M. Matijaš, J. A. Suykens, and S. Krajcar, "Load forecasting using a multivariate meta-learning system," *Expert systems with applications*, vol. 40, no. 11, pp. 4427-4437, Sep. 2013.

25. A. Arjmand, R. Samizadeh, and M. D, Saryazdi, "Meta-learning in multivariate load demand forecasting with exogenous meta-features," *Energy Efficiency*, pp. 1-17, Feb. 2020.

26. X. Wang, K. Smith-Miles, and R. Hyndman, "Rule induction for forecasting method selection: Meta-learning the characteristics of univariate time series," *Neurocomputing*, vol. 72, no. 10-12, pp. 2581-2594, Jun. 2009.

27. T. S. Talagala, R. J. Hyndman, and G. Athanasopoulos, "Meta-learning how to forecast time series," *Monash Econometrics and Business Statistics Working Papers*, vol. 6, pp.18, Apr. 2018.

28. C. Lemke, and B. Gabrys, "Meta-learning for time series forecasting and forecast combination," *Neurocomputing*, Vol. 73, no. 10-12, pp. 2006-2016, Jun. 2010.

29. J. Hu, J. Heng, J. Tang, and M. Guo, "Research and application of a hybrid model based on Meta learning strategy for wind power deterministic and probabilistic forecasting," *Energy Conversion and Management*, vol. 173, pp. 197-209, Oct. 2018.

30. J. R. Rice, "The algorithm selection problem," In *Advances in computers*, vol. 15, Elsevier, 1976, pp. 65-118.

31. T. Hospedales, A. Antoniou, P. Micaelli, and A. Storkey, "Meta-learning in neural networks: A survey," *arXiv preprint*, arXiv:2004.05439, 2020.

32. S. Ali, and K. A. Smith-Miles, "A meta-learning approach to automatic kernel selection for support vector machines". *Neurocomputing*, vol. 70, no. 1-3, pp. 173-186. Dec. 2006.

33. C. M., Lee, and C. N. Ko, "Short-term load forecasting using lifting scheme and ARIMA models," *Expert Systems with Applications*, vol. 38, no. 5, pp. 5902-5911, May, 2011.

34. T. Fang, and R. Lahdelma, "Evaluation of a multiple linear regression model and SARIMA model in forecasting heat demand for district heating system," *Applied energy*, vol. 179, pp. 544-552. Oct. 2016.

35. W. Kong, Z. Dong, D. J. Hill, F. Luo, and Y. Xu, "Short-term residential load forecasting based on resident behavior learning," *IEEE Trans. Power Syst.*, vol. 33, no. 1, pp. 1087-1088, Mar. 2017.

36. Y. Chen, P. Xu, Y. Chu, W. Li, Y. Wu, L. Ni, Y. Bao, and K. Wang, "Short-term electrical load forecasting using the Support Vector Regression (SVR) model to calculate the demand response baseline for office buildings," *Applied Energy*, vol. 195, pp. 659-670, 2017.

37. Y. Chen, P. B. Luh, C. Guan, Y. Zhao, L. D. Michel, M. A. Coolbeth, P. B. Friedland, and S. J. Rourke, "Short-term load forecasting: Similar day-based wavelet neural networks," *IEEE Trans. Power Syst.*, vol. 25, no. 1, pp. 322-330, Nov. 2009.

38. J. Benesty, J. Chen, Y. Huang, and I. Cohen, "Pearson correlation coefficient," In *Noise reduction in speech processing*, Springer, Berlin, Heidelberg, 2009, pp. 1-4.

39. A. Lahouar, and J. B. H. Slama, "Day-ahead load forecast using random forest and expert input selection," Energy Conversion and Management, vol. 103, pp. 1040-1051, Oct. 2015.

40. W. Gao, S. Oh, and P. Viswanath, "Demystifying Fixed k-Nearest Neighbor Information Estimators," *IEEE Trans. Info. Theory*, vol. 64, no, 8, pp. 5629-5661, Feb. 2018.

41. X. Z. Wang, Y. L. He, and D. D. Wang, "Non-naive Bayesian classifiers for classification problems with continuous attributes," *IEEE Trans. Cyber.*, vol. 44, no. 1, pp. 21-39, Feb. 2013.

42. A. Tharwat, "Linear vs. quadratic discriminant analysis classifier: a tutorial," *International Journal of Applied Pattern Recognition*, vol. 3, no. 2, pp. 145-180, Sep. 2016.

43. Pecan Street Dataport, 2020. [Online]. Available:   https://www. pecanstreet.org/dataport/

44. National Oceanic and Atmospheric Administration, 2020. [Online]. Available: https://www.noaa.gov/

45. Q. Zhou, M. Shahidehpour, A. Paaso, S. Bahramirad, A. Abdulwhab, and A. M. Abusorrah, "Distributed Control and Communication Strategies in Networked Microgrids," *IEEE Commun. Surveys Tuts.*, early access, 2020, doi: 10.1109/COMST.2020.3023963.

46. L. V. D. Maaten, and G. Hinton, "Visualizing data using t-SNE," *Journal of machine learning research*, vol. 9, pp. 2579-2605, Nov. 2008.